\RequirePackage[loading]{tracefnt}
\RequirePackage[loading]{tracefnt}
\documentclass[fleqn,usenatbib,useAMS]{mnras}

\usepackage[T1]{fontenc}
\usepackage{ae,aecompl}
\usepackage{newtxtext,newtxmath}

\usepackage{graphicx}	
\usepackage{amsmath}	
\usepackage{amssymb}	
\usepackage{multicol}        
\usepackage{bm}		
\usepackage{pdflscape}	

\title[Fate of the gaseous disks of galaxies in clusters]{The fate of the gaseous disks of galaxies that fall into clusters}
\author[Ruggiero \& Lima Neto]{Rafael Ruggiero\thanks{E-mail: rafael.ruggiero at usp.br} and Gastao B. Lima Neto
\\
Instituto de Astronomia, Geof\'isica e Ci\^encias Atmosf\'ericas, Universidade de S\~ao Paulo, R. do Mat\~ao 1226, 05508-090 S\~ao Paulo, Brazil}
\date{Accepted 2017 March 23. Received 2017 March 21; in original form 2016 December 12} 
\pubyear{2016}

\begin{document}
\label{firstpage}
\pagerange{\pageref{firstpage}--\pageref{lastpage}}
\maketitle

\begin{abstract}
Galaxy clusters are known to induce gas loss in infalling galaxies due to the ram pressure exerted by the intracluster medium over their gas content. In this paper, we investigate this process through a set of simulations of Milky Way like galaxies falling inside idealised clusters of 10$^{14}$ M$_\odot$ and 10$^{15}$ M$_\odot$, containing a cool-core or not, using the adaptive mesh refinement code RAMSES. We use these simulations to constrain how much of the initial mass contained in the gaseous disk of the galaxy will be converted into stars and how much of it will be lost, after a single crossing of the entire cluster. We find that, if the galaxy reaches the central region of a cool-core cluster, it is expected to lose all its gas, independently of its entry conditions and of the cluster's mass. On the other hand, it is expected to never lose all its gas after crossing a cluster without a cool-core just once. Before reaching the centre of the cluster, the SFR of the galaxy is always enhanced, by a factor of 1.5 to 3. If the galaxy crosses the cluster without being completely stripped, its final amount of gas is on average two times smaller after crossing the 10$^{15}$ M$_\odot$ cluster, relative to the 10$^{14}$ M$_\odot$ cluster. This is reflected in the final SFR of the galaxy, which is also two times smaller in the former, ranging from 0.5 -- 1 M$_\odot$ yr$^{-1}$, compared to 1 -- 2 M$_\odot$ yr$^{-1}$ for the latter.
\end{abstract}


\section{Introduction}
Galaxy clusters are among the highest density environments in which galaxies in the universe can evolve. Both star formation and nuclear activity depend on the local density \citep[e.g.][]{2004MNRAS.353..713K}, with the inner region of clusters representing the low end of both. Galaxies in clusters are also morphologically different from those in lower density environments, being predominantly early-type \citep{1980ApJ...236..351D}. Both these factors are related to the higher fraction of red galaxies in galaxy clusters \citep{2006MNRAS.373..469B}. Clearly, understanding the effect of clusters in the evolution of galaxies is a key ingredient in understanding how galaxies in the universe evolve \citep*[for a review on the effect of environment on galaxy evolution, see][]{2006PASP..118..517B}.

Of particular interest is the study of the interaction between gas-rich, disk galaxies and the intracluster medium (ICM) of galaxy clusters. The ram pressure exerted by the ICM removes part of the gas in the disk of the galaxy, starting from its less gravitationally bound outer regions \citep{1972ApJ...176....1G}. This process is called ram pressure stripping (RPS), and radio/optical observations of nearby galaxy clusters show evidence that it results in galaxies with truncated H$_\mathrm{I}$ disks embedded in untruncated stellar disks -- for instance in Virgo \citep{2004ApJ...613..866K,2008AJ....136.1623C} and in Coma \citep{2015AJ....150...59K}. A timescale of $\lesssim$ 1 Gyr is required for the RPS to remove the galaxy from the blue cloud, but at least a few Gyr are necessary for it to reach the red sequence as a result of the quenching of star formation \citep{2009MNRAS.400.1225C}. However, the hot gaseous halo of the galaxy is removed within a much shorter timescale, of $\sim$100 Myr \citep{2016A&A...591A..51S}.

The gradual removal of gas associated with moderate ram pressure cases may be the most relevant mechanism for morphological change in galaxy clusters \citep{2010ApJ...714.1779V}. However, some galaxies experience more extreme ram pressure scenarios, due to a combination of high velocities relative to the ICM and high ICM densities, causing a violent stripping of gas from theirs disks. This is often associated with ``jellyfish'' morphologies, in which the galaxy is observed with a system of gaseous clumps and filaments emerging from the disk and forming a tail behind it. Recently hundreds of such ``jellyfish galaxies'' have been discovered by applying a machine learning algorithm to Hubble images \citep{2016MNRAS.455.2994M}. With this larger statistical sample, it has been possible to infer that these extreme ram pressure cases might be preferably found in merging clusters, and not in virialised clusters gradually accreting galaxies from filaments. The vigorous star formation taking place in the compressed gas of some jellyfish galaxies temporarily turns them in some of the most luminous galaxies in the entire cluster \citep{2014ApJ...781L..40E}.

The processing of gas by ram pressure gives rise to a population of galaxies which have distinct features relative to the ones which haven't been near the cluster centre yet. These galaxies are nicknamed ``backsplash galaxies'' \citep[e.g.][]{2006MNRAS.366..645P,2011MNRAS.411.2637P}, and they systematically are redder have older stellar populations than galaxies which are infalling for the first time in the cluster \citep{2014A&A...564A..85M}. Numerical simulations have shown that half the galaxies which currently reside between 1 -- 2 virial radii of the host galaxy cluster are backsplash galaxies \citep{2005MNRAS.356.1327G}, evidence that they represent a fundamental component of the galaxy population in clusters.

Many numerical simulations have been run to investigate in detail the process of ram pressure stripping. These simulations have considered its role in generating S0 galaxies in clusters \citep{1980ApJ...241..928F,1999MNRAS.308..947A}, showing that they could be ram pressure stripped spiral galaxies \citep[][although, for example, \citet{2016MNRAS.460.1059J} argue that multiple tidal interactions between falling spirals and cluster members may still be required to form low specific angular momentum galaxies such as S0s]{2000Sci...288.1617Q}. They have also shown that the rate of gas loss weakly depends on the inclination angle of the disk \citep{2006MNRAS.369..567R}; that the star formation rate of the galaxy can increase by up to a factor of 3 because of the compression provided by the ram pressure \citep{2008A&A...481..337K}, although this increase can be smaller if a stellar bulge is present \citep{2012A&A...544A..54S}; that the tail morphology, but not the rate of gas loss, is affected by both the presence of magnetic fields \citep{2014ApJ...784...75R,2014ApJ...795..148T} and viscosity \citep{2008MNRAS.388L..89R}; and that the mean ram pressure experienced by galaxies within the virial radius of a cluster increases with redshift \citep{2010MNRAS.408.2008T}. Cosmological simulations of galaxy cluster formation have also shown that RPS happens for small cluster radii (< 1 Mpc) and on a timescale of more than 1 Gyr \citep{2007ApJ...671.1434T}; that for a given cluster radius, the ram pressure experienced by galaxies can change by an order of magnitude, in equal parts because of ICM substructure and of different relative velocities between the galaxy and the surrounding gas \citep{2008ApJ...684L...9T}; and that the ram pressure experienced by galaxies infalling from filaments may be up to two orders of magnitude larger \citep{2013MNRAS.430.3017B}. In terms of metal enrichment of the ICM, RPS has been shown to relevant essentially at the core of the cluster, and little at its periphery \citep{2006A&A...452..795D}.

Most ram pressure simulations in the literature involve submitting a galaxy to uniform winds, without taking into account changes in wind velocity, density and temperature as the galaxy orbits the cluster. Although these simulations establish and characterise the possible regimes of ram pressure stripping, predicting the state of a galaxy after it crosses a real galaxy cluster based on them is not straightforward. A way to self consistently take these changes into account is to include a whole idealised cluster in the simulation, in which the galaxy falls. Not many simulations like that have ever been run \citep[some examples are][]{2007MNRAS.380.1399R,2014MNRAS.438..444B,2016A&A...591A..51S}.

Our goal in this paper is to create a comprehensive picture for what changes a disk galaxy is expected to undergo after it crosses a realistic galaxy cluster. For this, we setup a grid of simulations in which a Milky Way like galaxy falls radially into four different, representative galaxy clusters: the total mass is either $10^{14}$ M$_\odot$ or $10^{15}$ M$_\odot$, and the density profile for the ICM either contains a cool-core or not. In order to also make the entry conditions for the galaxy representative, and thus cover a wide range of ram pressure intensities, we include three different entry speeds (0.5$\sigma$, $\sigma$ and 2$\sigma$, where $\sigma$ is the velocity dispersion of the cluster), and three different orientation angles between the normal to the disk's plane and its direction of motion (0$^\circ$, 45$^\circ$ and 90$^\circ$). This choice of free parameters leads to $36$ different scenarios -- significantly more than what previous models which included a whole cluster have covered. With these simulations, we predict how much of the initial gas mass of the galaxy can be turned into stars and how much can be lost as a function of the free parameters, and well as how its star formation rate will change over time.

This paper is structured as follows. In Section 2 we describe the methodology and the setup of our simulations; in Section 3 we describe the results, focusing on what happens to the initial gas disk after the cluster is crossed in each scenario; and in Section 4 we discuss our results in terms of both previous numerical models and observational results in the literature, and summarise.

\section{Simulations}

In order to constrain the transformations a disk galaxy will undergo after falls into a galaxy cluster and crosses it, we setup simulations of a Milky Way like galaxy falling radially into a set of idealised clusters in hydrostatic equilibrium, starting from a set of different entry conditions. 

\subsection{Numerical methods}

Our simulations use the Adaptive Mesh Refinement (AMR) code RAMSES \citep{2002A&A...385..337T}. This code solves the Euler equations for hydrodynamics in a dynamically refined cubic mesh using a second order Godunov scheme, and in the case of our simulations by applying the HLLC Riemann solver. 

We have introduced two modifications in the code for the purposes of this study. The first is the introduction of an external potential in the simulation box, representing the dark matter halo of the galaxy cluster, which as we will describe in Section \ref{sec:initcond} is the only component of our simulations represented externally. The second is the introduction of a cooling switch in the central region of the box, defined as a sphere with a user-provided radius, with the purpose of only allowing the cluster's core to cool in the short time window of the simulation when the galaxy is actually within it, thus preventing the high cooling rate of this region from causing a cooling-flow. Cooling is always enabled outside this region, and inside it is disabled both at the beginning and at the end of the simulation. It is only enabled there for $t_{\mathrm{cool},1} < t < t_{\mathrm{cool},2}$, where $t_{\mathrm{cool},1}$ and $t_{\mathrm{cool},2}$ are user-provided times. For our simulations, we calculate these times by integrating the galaxy's orbit in the total cluster potential numerically, and by picking the time when the centre of the galaxy is 20 kpc from reaching the central region and the time when it is 20 kpc past it, to ensure that cooling is always enabled in the surroundings of its disk. The way we define the radius of the central region is described in Section \ref{subs:paramspace}.

Our simulation box has a side length of 16 Mpc, and is refined up to 16 levels of refinement, resulting in a maximum resolution of 244 pc. The boundary conditions are periodic, and the box has been verified to be large enough for no apparent boundary effects to be taking place in it. In order to assess the numerical robustness of our results, we also include one simulation with an additional level of refinement, resulting in a maximum resolution of 122 pc.

We chose to employ two refinement criteria at once for the grid cells in our simulations. A cell is refined if it contains more than 80 particles, or if its size exceeds the local jeans length divided by 173. The former ensures that the dark matter halo of the galaxy will be resolved, and that its stellar disk will be resolved even if all its gas is lost. This is possible because the criterion considers both the star particles in the galaxy (which include its stellar disk and bulge, described in Section \ref{sec:initcond}) and its dark matter halo particles. The criterion applies specifically to the galaxy, as the cluster doesn't have any live collisionless component. The latter was empirically found to force all the gaseous disk of the galaxy to be refined up to the maximum level of refinement, while keeping the ICM in equilibrium, but not as heavily resolved as it would if a simple mass-based criterion was used, thus greatly saving computational resources. 

Radiative cooling is included in the simulations assuming a constant metallicity of $1$ $Z_\odot$ throughout the whole box. This value is representative of the Milky Way's ISM \citep{2014A&A...565A..89B}, and since by design the dynamics of the ICM in our clusters is not affected by cooling, the exact value of the metallicity outside the gaseous disk of our galaxy is not relevant. Recipes for star formation and feedback by supernovae are also included. Star formation happens in a cell if its density exceeds a threshold of $\rho_0 = 0.1$ H/cm$^3$, with the star formation rate $\dot{\rho_\star}$ given by a Schmidt law \citep{1959ApJ...129..243S}:

\begin{equation}
\dot{\rho_\star} = \epsilon \frac{\rho_{\mathrm{gas}}}{t_\mathrm{ff}}, \qquad\text{where } t_\mathrm{ff} \text{ is the local free-fall time}.
\end{equation}

We choose a value of 1\% for the star formation efficiency per free-fall time $\epsilon$, which is similar to values derived from observations of young stellar objects in the Milky Way \citep{2007ApJ...654..304K}. In the cells which are flagged for star formation, the equation of state ceases to be that of an ideal gas and becomes a polytrope, in order to account for sub-grid thermal and turbulent motions in the ISM which would otherwise prevent the temperature of the gas from dropping too much \citep{2011MNRAS.414..195T}:

\begin{equation}
T = T_0 \left(\frac{\rho_\mathrm{gas}}{\rho_0}\right)^{\kappa-1}.
\end{equation}
For the free parameters in this equation of state we pick $T_0$ = $10^4$ K \citep[as in][]{2011MNRAS.414..195T} and $\kappa = 2$ \citep[which is a widely used value, e.g.][]{2015ApJ...812L..36B}. These parameters combined with those of the star formation recipe lead to an initial SFR of $1.9$ M$_\odot$ yr$^{-1}$ for the galaxy when it is simulated in isolation, close to observational values derived for the Milky Way \citep{2010ApJ...710L..11R}. 

Feedback by supernovae is included using the thermal feedback recipe available in RAMSES, which takes into account the release of mass and energy by SNe II. This recipe depends on a free parameter $\eta$, which is the mass fraction of newly formed stars that explode as supernovae, and which we choose as 10\% following \citet{2013MNRAS.429.3068T}.

\subsection{Initial conditions} \label{sec:initcond}

The initial conditions for the isolated galaxy and isolated clusters in our simulations were generated using the codes galstep\footnote{\url{https://github.com/ruggiero/galstep}} and clustep\footnote{\url{https://github.com/ruggiero/clustep}}, respectively. Both these codes output the initial conditions as binary files in the entry format of the code GADGET-2 \citep{2005MNRAS.364.1105S}. These isolated initial conditions are concatenated considering the initial position of the galaxy, which is at the $R_{200}$ of the cluster in all cases, defined as the radius within which the average density of the cluster is equal to 200 times the critical density of the universe -- taken at the present time and using the Planck 2015 cosmological parameters \citep{2016A&A...594A..13P}; as well as its entry speed and the orientation of its disk, generating the initial conditions that are used in our simulations. These initial conditions in the GADGET-2 format are directly read into RAMSES using the DICE patch\footnote{\url{https://bitbucket.org/vperret/dice/wiki/RAMSES\%20simulation}} included in the code, which transfers the particles into the RAMSES particle tree and transfers the mass of the gas particles to the corresponding AMR cells using a Nearest Grid Point (NGP) scheme.

The galaxy consists of a dark matter halo, a stellar disk, a stellar bulge and a gaseous disk. A gaseous halo is not included because, in a RPS scenario, it is expected to be stripped in a short timescale \citep[$\sim$100 Myr,][]{2016A&A...591A..51S}, therefore not significantly affecting the star formation rate in the galaxy's disk. Both the dark matter and the bulge follow a Hernquist density profile \citep{1990ApJ...356..359H}:

\begin{equation}
\rho(r) = \frac{M}{2\pi} \frac{a}{r} \frac{1}{(r+a)^3},
\end{equation}
where $M$ is the total mass of the component and $a$ is a scale factor. This profile is similar to the NFW profile for dark matter halos \citep{1996ApJ...462..563N}, with the advantage that its total mass is finite. We chose $M = 10^{12}$ M$_\odot$ for the dark matter and $M = 10^{10}$ M$_\odot$ for the bulge. The value of $a$ is defined following the relation between $M_{200}$ (the mass within $R_{200}$) and concentration for dark matter halos of \citet{2008MNRAS.390L..64D}. For this we use the following procedure. For a given value of $a$, we calculate the $M_{200}$ associated with the Hernquist profile with $M = 10^{12}$ M$_\odot$ and this value of $a$. Then, we find the NFW profile associated with this $M_{200}$ using the \citet{2008MNRAS.390L..64D} relation. Finally, we compute distance between the two profiles, measured as the $\chi^2$ for a large sample of points with 0 Mpc < r $\leq$ 2.5 Mpc. The parameter $a$ is optimised to minimise this distance. This procedure results in a value of $a = 47$ kpc for the halo. For the bulge, we chose $a = 1.5$ kpc, corresponding to an effective radius of 3.6 kpc. 

Both disks follow an exponential density profile:

\begin{equation} \label{eq:rhodisk}
\rho(R, z) = \frac{M}{4 \pi R_d^2 z_0} \exp\left(-\frac{R}{R_d}\right) \mathrm{sech}^2\left(\frac{z}{z_0}\right).
\end{equation}
The mass of the stellar disk is $5 \times 10^{10}$ M$_\odot$, and the mass of the gaseous disk is 20\% of this value, $1\times 10^{10}$ M$_\odot$. The radial scale $R_d$ is the same for both disks, and is equal to 3.5 kpc. The vertical scale $z_0$ is 0.7 kpc for the stellar disk, and 2.5\% of this value for the gaseous disk. The initial temperature of the latter is $10^4$ K, which is a typical value for the warm ionised medium of the Milky Way \citep{2001RvMP...73.1031F}. The initial values for the vertical scale and the temperature of the gaseous disk are such that, right at the beginning of the simulation, the radiative cooling makes it settle with a vertical scale equal to twice the smallest cell size and a temperature of $\lesssim$ 10$^4$ K.

Velocities are assigned using the prescription found in \citet{2005MNRAS.361..776S}. One free parameter of this approach is the scale factor $f_R \equiv \sigma_R^2 / \sigma_z^2$, which defines the radial velocity dispersion of the disk at each point once its vertical velocity dispersion in this point has been calculated. We choose $f_R$ = 0.8 in our model, which results in a relatively stable disk when the galaxy is simulated in isolation.

The clusters contain a dark matter halo and a gaseous component, with 10\% of the total cluster mass in all cases -- which is a rather typical gas mass fraction for galaxy clusters with mass above $10^{14}$ M$_\odot$ \citep{2013A&A...555A..66L}. The dark matter halo is represented by an external potential introduced in RAMSES, which drastically reduces the number of particles in the simulation and thus its computational cost, while keeping the orbit of the galaxy realistic. The density profile for the dark matter is a Hernquist density profile, with the scale factor $a$ chosen using the same procedure we have described for the halo of the galaxy, using the \citet{2008MNRAS.390L..64D} relation. The gas follows a Dehnen density profile \citep{1993MNRAS.265..250D}:

\begin{equation}
\rho(r) = \frac{(3-\gamma)M}{4 \pi} \frac{a}{r^\gamma (r+a)^{4-\gamma}}.
\end{equation}
The value of $a$ for the gas is always equal to the value of $a$ of its dark matter halo, for simplicity. Temperatures are assigned to the gas to ensure hydrostatic equilibrium, and the ICM is assumed static -- no velocities are assigned to the gas particles. We consider two values for the parameter $\gamma$, namely $\gamma = 0$ and $\gamma = 1$. If $\gamma = 1$, then the density profile corresponds to a Hernquist density profile, which has a central cusp, and which features a drop in temperature in the central region. We associate this density profile with a ``cool-core'' type of cluster. If $\gamma = 0$, the density profile is flat in the central region and features no drop in temperature, qualitatively resembling a $\beta$-model \citep{1976A&A....49..137C} for the ICM. We associate this density profile with a cluster without a cool-core.

The mass resolution for the collisionless components of the simulations is $5\times10^{5}$ M$_\odot$. This results in $2\times10^6$ particles in the dark matter halo of the galaxy, $1\times10^5$ particles in its stellar disk and $4\times10^4$ particles in its bulge. The gas particles are only considered when the initial conditions are read, as their mass is transferred into the AMR grid cells. In order to get a good spatial coverage of the gas particles in the initial conditions and not leave many grid cells unfilled, we include a high number of gas particles in the initial conditions, namely $10^7$ in each galaxy cluster and $2\times10^6$ in the galaxy's disk.

\subsection{Parameter space} \label{subs:paramspace}

We chose to consider four free parameters for the interaction between the galaxy and the clusters in this study: the mass of the cluster, which is either 10$^{14}$ M$_\odot$ or 10$^{15}$ M$_\odot$; whether or not the density profile of the ICM exhibits a cool-core, which as described before we associate with $\gamma = 1$ (cool-core) or $\gamma = 0$ (no cool-core) in the Dehnen density profile assigned to the gas component; the entry speed of the galaxy, which is either $0.5\sigma$, $\sigma$ or $2 \sigma$, where $\sigma$ represents the velocity dispersion of the cluster, calculated as the standard deviation of the speeds of the dark matter particles contained within $R_{200}$; and the orientation of the disk of the galaxy, which is either 0$^\circ$ (face-on), 45$^\circ$ or 90$^\circ$ (edge-on). We note that the galaxy always falls radially, which is not the result of an assumption on the velocity distribution of the galaxies in the clusters we are modeling, but merely a way to explore a wide range of scenarios in which ram pressure is significant.

Table \ref{tab:clusters} lists the $R_{200}$, the velocity dispersion and the radius within which cooling is initially switched off for each cluster we are considering in this study. These radii are chosen by simulating each cluster in isolation for the time it takes the galaxy with the slowest initial speed ($0.5 \sigma$) to cross it, and by inspecting up to what radius the gas radial density profile was altered due to the cooling-flow. We note that no cooling-flow takes place in the cluster with $10^{15}$ M$_\odot$ and no cool-core during this time.

\begin{table}
 \caption{The properties of the clusters considered in this study.}
 \label{tab:clusters}
 \begin{tabular}{ccccccc}
Mass (M$_\odot$) & cool-core & $R_{200}$ (kpc) & $\sigma$ (km/s) & $r_{\mathrm{cool}}$ (kpc)\\
\hline
10$^{14}$ & yes & 805.1 & 352.5 & 100 \\
10$^{14}$ & no & 796.5 & 344.8 & 150 \\
10$^{15}$ & yes & 1679.0 & 713.2 & 50 \\
10$^{15}$ & no & 1657.9 & 697.0 & -- \\
 \end{tabular}
\end{table}

\section{Results}

In order to calculate quantities associated to the disk of the galaxy at each snapshot of the simulation, we considered a cylinder of radius 20 kpc and height 10 kpc placed at the centre of mass of its initial star particles, and oriented considering the entry angle of the galaxy in each simulation. This cylinder includes all cells belonging to the disk of the galaxy, but it also includes cells belonging to the ICM, and these two must be disentangled. Gas cells in the cylinder are assumed to belong to the galaxy if their temperature is below $10^6$ K -- which is a temperature that the dense disk of the galaxy can't reach, given its high cooling rate; and if their velocity along the direction of motion of the galaxy is within 5 standard deviations of the average of the velocities of the initial stellar disk in that direction. By analysing the velocity distributions of the galaxy, this has been verified to include all stars and gas cells in the disk, while excluding components with velocities close to 0, which would necessarily be a part of the ICM. Likewise, star particles in the cylinder for which this velocity criterion is satisfied are assumed to belong to the galaxy, thus disconsidering any eventual star formation happening at the ejected tail of the galaxy. The stars which belong to the disk of the galaxy itself remain attached to it during the whole simulation.

\begin{figure}
 \includegraphics[width=\columnwidth]{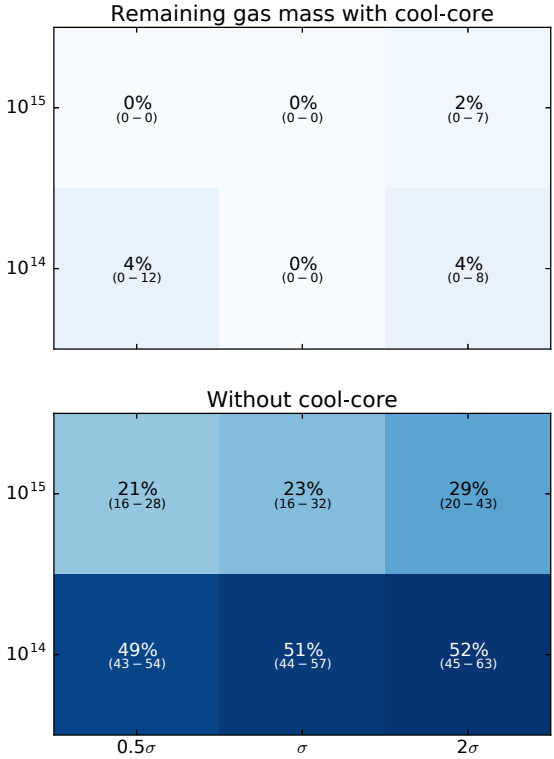}
 \caption{Gas mass in the galaxy after a single crossing of the cluster, as a function of velocity (x-axis), which is given in terms of the velocity dispersion of the cluster; and of cluster mass in M$_\odot$ (y-axis). The central values represent averages for the three entry angles, and the values in parenthesis the minimum and maximum values for these. Clusters with a cool-core typically leave the galaxy without any gas after a single crossing, contrary to what happens in a cluster without one. For the latter, the gas mass that remains in the galaxy decreases on average by a factor of two when the mass of the cluster goes from $10^{14}$ M$_\odot$ to $10^{15}$ M$_\odot$.}
 \label{fig:gas}
\end{figure}

We have chosen to marginalise the entry angle of the galaxy throughout all our analysis, as previous work have characterised its effect in details \citep[e.g.][]{2006MNRAS.369..567R}. Instead, we use the different entry angles to provide statistical significance to our results. Figure \ref{fig:gas} shows how much of the original gas mass of the galaxy remains attached to it after a single crossing of the cluster, considering only the gas in ISM form, and not the gas mass that was converted into stars. The central values displayed in each square are the averages for the three entry angles, and the values in parenthesis represent the minimum and maximum values for these. It can be noted that, regardless of entry conditions or cluster mass, little to no ISM is expected to remain in the galaxy if it crosses the central region of a cool-core cluster.

On the other hand, for the cluster without a cool-core, the galaxy is expected to never lose all its gas content after a single crossing of the cluster. As our simulations only include radial orbits, which among all possible orbits are the ones that reach the highest speeds and highest ICM densities (and hence highest ram pressure values), the values we present represent lower bounds for the remaining gas mass after a single crossing. Clearly, this gas mass is dependent on the cluster mass -- a $10^{15}$ M$_\odot$ cluster removes on average a factor of two more gas than a $10^{14}$ M$_\odot$ cluster.

\begin{figure}
 \includegraphics[width=\columnwidth]{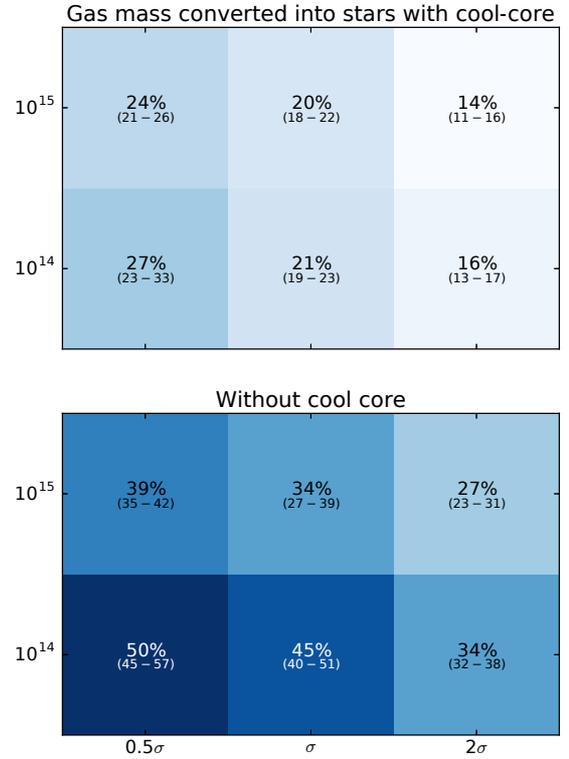}
 \caption{The same as Figure \ref{fig:gas}, but for how much of the initial gas mass in the galaxy is turned into stars after a single crossing of the cluster. Overall, more gas is converted into stars in clusters without a cool-core, because the ram pressure in these is less intense, leaving the galaxy with gas to form stars for longer.}
 \label{fig:stars}
\end{figure}

A similar analysis has been carried out to calculate how much of the initial gas mass in the galaxy is converted into stars after a crossing of the cluster, as displayed in Figure \ref{fig:stars}. The most prominent pattern is that a cluster with a cool-core is less efficient at forming stars than a cluster without one. The ram pressure stripping process is more efficient in the cool-core clusters, making the galaxy lose its gas faster, and consequently rendering it unable to keep its star formation steady over the entire orbit. Indeed, after the central passage, when the galaxy loses all its gas, the SFR evidently drops to zero in the cool-core cluster.

\begin{figure}
 \includegraphics[width=\columnwidth]{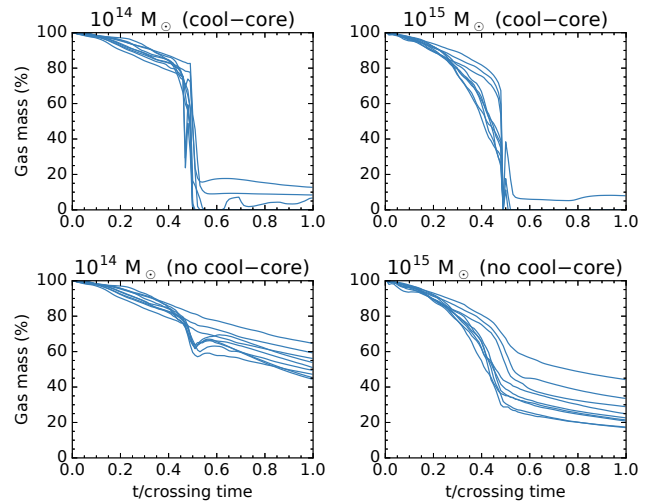}
 \caption{Gas mass in our galaxies as a function of time, normalised by how long it takes for each galaxy to cross the cluster. Initially the rate of gas loss is similar for all galaxies because it is dominated by star formation, but it increases more drastically over time for the cool-core clusters, resulting in a complete stripping of the gaseous disk.}
 \label{fig:gasmass}
\end{figure}

\begin{figure}
 \includegraphics[width=\columnwidth]{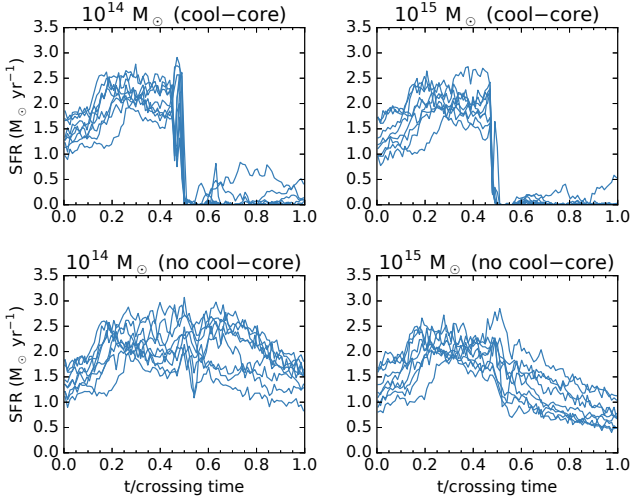}
 \caption{Similar to Figure \ref{fig:gasmass}, but for the star formation rate of the galaxies as a function of time. In the cool-core clusters, the SFR drops to zero after the central passage because all the gas is lost. The SFR initially increases by a factor of 1.5 -- 3 for all galaxies.}
 \label{fig:sfr}
\end{figure}

The global picture for the process of gas loss can be inferred from Figure \ref{fig:gasmass}, and is the following. In all scenarios, the galaxy will lose gas as it falls into the cluster, as a result of both star formation in the absence of a coronal supply of gas and of ram pressure stripping. In the very beginning of the orbit, when the galaxy is still in the outskirts of the cluster, the ram pressure is weak and the rate of gas loss is dominated by the star formation, hence the similar initial derivatives for all curves. As the galaxy approaches the centre of the clusters, the ram pressure increases more quickly for a cool-core cluster, and so does the rate of gas loss. As can be seen in Figure \ref{fig:sfr}, the SFR of the galaxy does not change too drastically in any case, so that the larger rate of gas loss in the cool-core cluster has to be caused by a larger rate of gas stripping, and not of star formation.

It can be noted from Figure \ref{fig:sfr} that all galaxies have their star formation rate enhanced initially, before reaching the centre of the clusters. The ratio between the maximum SFR of the galaxies before reaching the centre of the clusters and their initial SFR ranges from 1.5 to 3, with an average value of 2.0 that does not depend on cluster mass or density profile. For the galaxies which are not completely stripped, i.e. the ones that cross the cluster without a cool-core, the final SFR after having crossed the cluster range from 1 -- 2 M$_\odot$ yr$^{-1}$ for the 10$^{14}$ M$_\odot$ cluster, with an average of 1.5 M$_\odot$ yr$^{-1}$; and from 0.5 -- 1 M$_\odot$ yr$^{-1}$ for the 10$^{15}$ M$_\odot$ cluster, with an average of 0.7 M$_\odot$ yr$^{-1}$. This factor of two between the two star formation rates is expected, as the 10$^{15}$ M$_\odot$ cluster removes on average a factor of two more gas than the 10$^{14}$ M$_\odot$ cluster.

An illustrative view of the process of gas loss modeled in our simulations can be seen in Figure \ref{fig:panels}. All the displayed galaxies are crossing the $10^{14}$ M$_\odot$ cluster with an initial speed equal to $\sigma$. The top three ones are crossing the cool-core version cluster, and lose all their gas mass after the central passage. The lower three galaxies are crossing the version without a cool-core, and after crossing the cluster they have their gas radius truncated, but without losing all their gas mass. We note that the $10^{15}$ M$_\odot$ version of this plot is qualitatively equivalent to what we display here.

\subsection*{Numerical convergence}

The output of hydrodynamic simulations is subject to convergence issues if not enough resolution is used to follow the flow. In particular, numerical diffusion in the AMR method might be significant if the resolution is too low. Thus, in order to assess how robust our results are, we ran two tests. The first was to run one of our 36 simulations with an additional level of refinement, resulting in a resolution of 122 pc in the disk of the galaxy. For this test, we chose the simulation with a $10^{14}$ M$_\odot$ cluster without a cool-core, in which a galaxy falls face-on with a speed equal to $2 \sigma$. The highest entry speed was chosen because, if numerical diffusion is present, it will be strongest in this case, and the no cool-core cluster was chosen because in the cool-core version the galaxy would lose all its gas in the central passage regardless of resolution, rendering the test less informative.

\begin{figure}
 \includegraphics[width=\columnwidth]{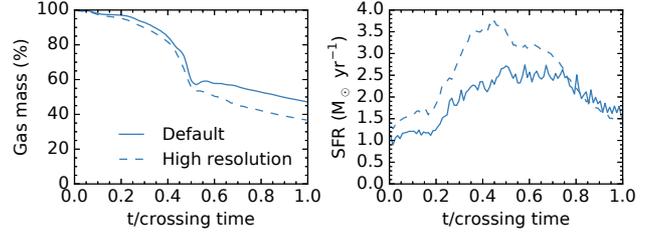}
 \caption{Test of the numerical convergence of our results. The galaxy falling face-on, with an initial speed of $2 \sigma$ (the highest one) into the $10^{14}$ M$_\odot$ cluster without a cool-core was resimulated with a two times higher resolution, reaching 122 pc. A larger initial increase in the SFR of the galaxy is noticed in the high resolution run. The amount of gas lost is also somewhat higher in this run, with a final difference of 10\% of the initial gas mass. The final SFR is similar in both cases.}
 \label{fig:convergence}
\end{figure}

The result of our test is summarised in Figure \ref{fig:convergence}. It can be noted that the initial star formation rate increase is more accentuated in the high resolution run, probably because this simulation is able to capture a higher compression of the gaseous disk. The final amount of gas in the high resolution run is smaller, mainly because more gas is lost in this case during the central passage. This happens because the stripped gas fragments in smaller clouds in the high resolution run, which are more efficiently mixed with the ICM and don't manage to leak back into the galaxy, like some larger clouds present in the standard resolution run do. The infall of these larger clouds causes an increase in the gas content of the standard resolution galaxy shortly after the central passage, as observed in left panel of Figure \ref{fig:convergence}, which is when the difference between the two runs arises. However, the final difference is of only 10\% of the initial gas mass in the disk, indicating that our results are at most mildly overestimating the final amount of gas in the galaxies. The final SFR is very similar in both runs.

The second test we considered was to place the galaxy stopped at the centre of the same cluster as in the previous test, and let it evolve without star formation or radiative cooling for the same time as the crossing time of the galaxy in that test, in order to assess how much gas will be lost by numerical diffusion alone. Initially the potential of the cluster destabilises the galaxy, but after that it settles in a state of equilibrium. The rate of gas loss of the settled galaxy is of $0.04$ M$_\odot$ yr$^{-1}$, which over one crossing time would result in only 0.7\% of its gas content being lost. Thus, we rule out numerical diffusion as a relevant cause of gas loss in our simulations.

\section{Discussion and summary}

\begin{figure*}[p]
 \includegraphics[width=\textwidth]{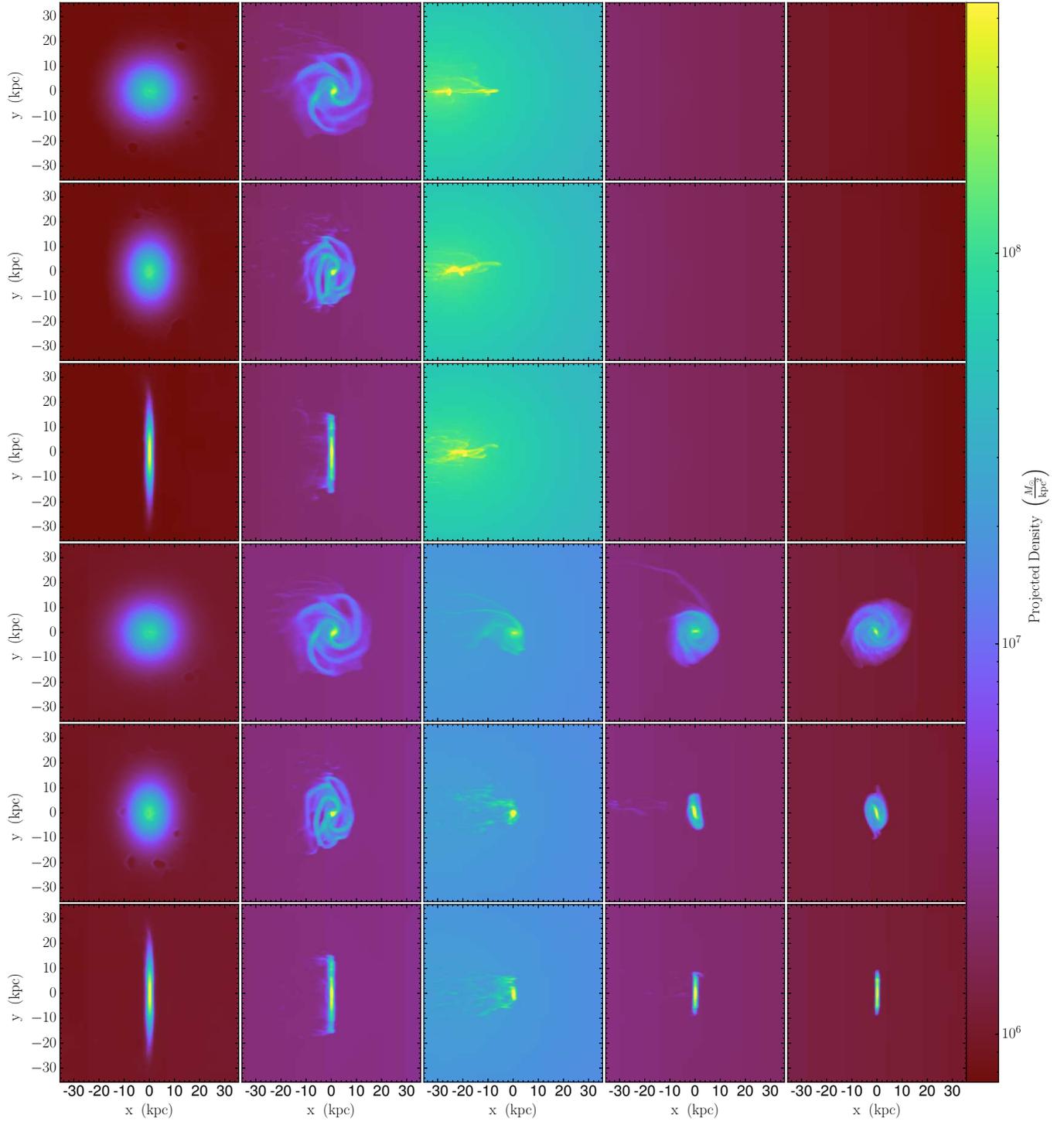}
 \caption{Projection plots showing the process of gas loss for six of our simulations. All galaxies are moving from left to right, and the columns correspond to 0\%, 25\%, 50\%, 75\% and 100\% of the crossing time for each case. In all cases the entry speed is equal to $\sigma$ and the cluster mass is $10^{14}$ M$_\odot$. The behaviour of the galaxies is qualitatively similar in the case of a $10^{15}$ M$_\odot$ cluster. First three rows: cluster with a cool-core, and galaxy entering it edge-on, at 45$^\circ$ and face-on. All the gas is lost after the central passage in all cases. Last three rows: the same for a cluster without a cool-core. The gaseous disk of the galaxies end up truncated.}
 \label{fig:panels}
\end{figure*}

Qualitatively, the changes in gas morphology we find in our simulations are within what would be expected, considering observations of real clusters \citep[e.g.][]{2004ApJ...613..866K} and classical models \citep{1972ApJ...176....1G} -- the gaseous disk either has its radius truncated or is completely stripped, depending on the ram pressure encountered by the galaxy along its orbit. One illustrative example is the galaxy in the last row of Figure \ref{fig:panels}. It is falling face-on, so that the stripping criterion of \citet{1972ApJ...176....1G} may be applied. The criterion states that the ram pressure must be smaller than the gravitational restoring force per unit area of the disk of the galaxy:

\begin{equation} \label{eq:gunngott}
\rho_\mathrm{ICM} v^2 < 2 \pi G \,\Sigma_\mathrm{gas} \Sigma_\mathrm{stars},
\end{equation}
where $\Sigma$ is the projected density, in our case given by the integral of Equation \ref{eq:rhodisk} for all values of z. For this galaxy, one finds that in the central passage its gaseous disk is expected to be truncated at a radius of $4.1$ kpc. In the central panel of Figure \ref{fig:panels}, the galaxy has a radius of $\sim$3 kpc, close to the predicted value. It is not surprising that this radius is smaller than predicted, as the galaxy has gas converted into stars before reaching the centre of the cluster. If the gas mass that was converted into stars is $m$, then, considering Equation \ref{eq:rhodisk},

\begin{equation} \label{eq:product}
\Sigma_\mathrm{gas} \Sigma_\mathrm{stars} \propto (M_\mathrm{gas,0} - m) (M_\mathrm{stars, 0} + m),
\end{equation}
where the subscript 0 denotes the initial values. If $M_\mathrm{stars,0} > M_\mathrm{gas,0}$, which is the case for our galaxy and for late-type galaxies in general, then the product in Equation \ref{eq:product} is maximum for $m = 0$, therefore rendering the \citet{1972ApJ...176....1G} prediction an upper bound for the stripping radius.

In the following two panels, this galaxy has the radius of its gas disk noticeably increased. This is because even if the ram pressure manages to push the gas away from the disk in the central passage, this gas may still be bound to the total potential well of the galaxy just after the central passage, when the ram pressure it experiences suddenly decreases. This is analogous to the backfall \citet{2007MNRAS.380.1399R} have observed in their simulations. 

The criterion in Equation \ref{eq:gunngott} explains why the cool-core clusters are capable of removing all the gas from the galaxy in just one crossing, while the clusters without a cool-core are not. The gas density in the central region of the former clusters considered in this study is 2 -- 3 orders of magnitude higher than in the latter, and consequently so is the ram pressure experienced by the galaxy. One particular observational example which can possibly be better understood by this result is the Virgo cluster, which generally seems to not be capable of completely stripping its member galaxies \citep{2008AJ....136.1623C}. As the Virgo cluster does not contain a cool-core, this could be one of the factors at play that prevented its galaxies from having been more stripped more efficiently. Of course this would be one of many possible factors, as e.g. ICM substructure away from the centre of the cluster can also be relevant \citep{2008ApJ...684L...9T}.

We emphasise that all the orbits considered in our model are radial, while in reality galaxies move inside of galaxy clusters with impact parameters different from zero. But our model can also be used to understand non-radial cases, as a real galaxy falling inside a cool-core galaxy cluster with a non-zero impact parameter will encounter a density profile similar to what our galaxies falling into a cluster without a cool-core find -- with an increase in density that doesn't reach the central peak. In terms of global properties, statistically some of the galaxies in a cluster will have orbits that penetrate deep into its potential well (e.g. by infalling with a small impact parameter), and these galaxies will go through a more drastic ram pressure stripping event in case they encounter a cool-core at the centre of the cluster. This population of galaxies will cause a difference in the averaged properties of the galaxies in clusters with and without a cool-core.

In terms of star formation, the results found in our simulations are similar to what previous models in the literature have found for the SFR, such as \citet{1999ApJ...516..619F}, who show that ram pressure can increase the SFR by up to a factor of two, and \citet{2008A&A...481..337K}, who show that it can increase by up to factor of three. The timescale for the star formation to cease we find is also consistent with previous observational results, like \citet{2009MNRAS.400.1225C}, who infer a timescale of at least a few Gyr. Our galaxies take typically 1 -- 2 Gyr to cross the clusters, and unless a galaxy falls radially into a cool-core cluster, after this time it will still be forming stars, as our results for the cluster without a cool-core show.

Based on our results on star formation, we infer that backsplash populations are sensitive to the presence of substructure in the galaxy cluster, and should feature consistently lower star formation rates in cool-core galaxy clusters, relative to clusters without a cool-core. One way to test this hypothesis would be to compare the backsplash populations in a sample of cool-core galaxy clusters stacked together with the populations in a stacked sample of clusters without a cool-core, assuming both samples are equivalent in terms of other variables which might affect the ram pressure efficiency, such as the dynamical state of the cluster, which might result in systematically larger relative velocities between the galaxy and the ICM, and thus larger ram pressure values.

What we find in the model we present here can be summarised as follows. Clusters with a cool-core are capable of stripping gas much more efficiently than clusters without one. Indeed, a galaxy that falls radially into a cool-core cluster is expected to lose all its gas mass after a single crossing of the cluster, while the galaxy will necessarily have to cross a cluster without a cool-core more than once to lose all its gas. Star formation is always initially increased due to ram pressure, by a factor of 1.5 to 3, due to the compression of the gaseous disk that the ram pressure causes. But this star formation may end sooner in cool-core clusters, because the gas of the galaxy can be removed earlier in these clusters. This makes clusters with a cool-core environments potentially more efficient at quenching star formation than clusters without one. Galaxies that cross a $10^{15}$ M$_\odot$ cluster in orbits that don't remove all their gas end up with a SFR of 0.5 -- 1 M$_\odot$ yr$^{-1}$, which is on average two times smaller than after crossing a $10^{14}$ M$_\odot$ cluster, where they end up with a SFR of 1 -- 2 M$_\odot$ yr$^{-1}$. This factor of two is a consequence of the fact that the $10^{15}$ M$_\odot$ cluster removes on average two times more gas from the galaxy than the $10^{14}$ M$_\odot$ cluster.

\section*{Acknowledgments}

This work has made use of the computing facilities of the Laboratory of Astroinformatics (IAG/USP, NAT/Unicsul), which purchase was made possible by the Brazilian agency FAPESP (grant 2009/54006-4) and the INCT-A. In particular, we used the 2304 cores cluster Alphacrucis. The research project of which this work is a part is funded by FAPESP (grant 2015/13141-7). GBLN thanks CNPq for partial financial support. All the analysis and the visualisation of our data were based on the Python package yt\footnote{\url{http://yt-project.org/}}, described in \citet{2011ApJS..192....9T}. We thank the anonymous reviewer for the constructive feedback.

\bibliographystyle{mnras}
\bibliography{references}

\begin{thebibliography}{}
\makeatletter
\relax
\def\mn@urlcharsother{\let\do\@makeother \do\$\do\&\do\#\do\^\do\_\do\%\do\~}
\def\mn@doi{\begingroup\mn@urlcharsother \@ifnextchar [ {\mn@doi@}
  {\mn@doi@[]}}
\def\mn@doi@[#1]#2{\def\@tempa{#1}\ifx\@tempa\@empty \href
  {http://dx.doi.org/#2} {doi:#2}\else \href {http://dx.doi.org/#2} {#1}\fi
  \endgroup}
\def\mn@eprint#1#2{\mn@eprint@#1:#2::\@nil}
\def\mn@eprint@arXiv#1{\href {http://arxiv.org/abs/#1} {{\tt arXiv:#1}}}
\def\mn@eprint@dblp#1{\href {http://dblp.uni-trier.de/rec/bibtex/#1.xml}
  {dblp:#1}}
\def\mn@eprint@#1:#2:#3:#4\@nil{\def\@tempa {#1}\def\@tempb {#2}\def\@tempc
  {#3}\ifx \@tempc \@empty \let \@tempc \@tempb \let \@tempb \@tempa \fi \ifx
  \@tempb \@empty \def\@tempb {arXiv}\fi \@ifundefined
  {mn@eprint@\@tempb}{\@tempb:\@tempc}{\expandafter \expandafter \csname
  mn@eprint@\@tempb\endcsname \expandafter{\@tempc}}}

\bibitem[\protect\citeauthoryear{{Abadi}, {Moore}  \& {Bower}}{{Abadi}
  et~al.}{1999}]{1999MNRAS.308..947A}
{Abadi} M.~G.,  {Moore} B.,   {Bower} R.~G.,  1999, \mn@doi [\mnras]
  {10.1046/j.1365-8711.1999.02715.x}, \href
  {http://adsabs.harvard.edu/abs/1999MNRAS.308..947A} {308, 947}

\bibitem[\protect\citeauthoryear{{Bah{\'e}}, {McCarthy}, {Balogh}  \&
  {Font}}{{Bah{\'e}} et~al.}{2013}]{2013MNRAS.430.3017B}
{Bah{\'e}} Y.~M.,  {McCarthy} I.~G.,  {Balogh} M.~L.,   {Font} A.~S.,  2013,
  \mn@doi [\mnras] {10.1093/mnras/stt109}, \href
  {http://adsabs.harvard.edu/abs/2013MNRAS.430.3017B} {430, 3017}

\bibitem[\protect\citeauthoryear{{Baldry}, {Balogh}, {Bower}, {Glazebrook},
  {Nichol}, {Bamford}  \& {Budavari}}{{Baldry}
  et~al.}{2006}]{2006MNRAS.373..469B}
{Baldry} I.~K.,  {Balogh} M.~L.,  {Bower} R.~G.,  {Glazebrook} K.,  {Nichol}
  R.~C.,  {Bamford} S.~P.,   {Budavari} T.,  2006, \mn@doi [\mnras]
  {10.1111/j.1365-2966.2006.11081.x}, \href
  {http://adsabs.harvard.edu/abs/2006MNRAS.373..469B} {373, 469}

\bibitem[\protect\citeauthoryear{{Bekki}}{{Bekki}}{2014}]{2014MNRAS.438..444B}
{Bekki} K.,  2014, \mn@doi [\mnras] {10.1093/mnras/stt2216}, \href
  {http://adsabs.harvard.edu/abs/2014MNRAS.438..444B} {438, 444}

\bibitem[\protect\citeauthoryear{{Bergemann} et~al.,}{{Bergemann}
  et~al.}{2014}]{2014A&A...565A..89B}
{Bergemann} M.,  et~al., 2014, \mn@doi [\aap] {10.1051/0004-6361/201423456},
  \href {http://adsabs.harvard.edu/abs/2014A%26A...565A..89B} {565, A89}

\bibitem[\protect\citeauthoryear{{Bieri}, {Dubois}, {Silk}  \& {Mamon}}{{Bieri}
  et~al.}{2015}]{2015ApJ...812L..36B}
{Bieri} R.,  {Dubois} Y.,  {Silk} J.,   {Mamon} G.~A.,  2015, \mn@doi [\apjl]
  {10.1088/2041-8205/812/2/L36}, \href
  {http://adsabs.harvard.edu/abs/2015ApJ...812L..36B} {812, L36}

\bibitem[\protect\citeauthoryear{{Boselli} \& {Gavazzi}}{{Boselli} \&
  {Gavazzi}}{2006}]{2006PASP..118..517B}
{Boselli} A.,  {Gavazzi} G.,  2006, \mn@doi [\pasp] {10.1086/500691}, \href
  {http://adsabs.harvard.edu/abs/2006PASP..118..517B} {118, 517}

\bibitem[\protect\citeauthoryear{{Cavaliere} \& {Fusco-Femiano}}{{Cavaliere} \&
  {Fusco-Femiano}}{1976}]{1976A&A....49..137C}
{Cavaliere} A.,  {Fusco-Femiano} R.,  1976, \aap, \href
  {http://adsabs.harvard.edu/abs/1976A%26A....49..137C} {49, 137}

\bibitem[\protect\citeauthoryear{{Cortese} \& {Hughes}}{{Cortese} \&
  {Hughes}}{2009}]{2009MNRAS.400.1225C}
{Cortese} L.,  {Hughes} T.~M.,  2009, \mn@doi [\mnras]
  {10.1111/j.1365-2966.2009.15548.x}, \href
  {http://adsabs.harvard.edu/abs/2009MNRAS.400.1225C} {400, 1225}

\bibitem[\protect\citeauthoryear{{Crowl} \& {Kenney}}{{Crowl} \&
  {Kenney}}{2008}]{2008AJ....136.1623C}
{Crowl} H.~H.,  {Kenney} J.~D.~P.,  2008, \mn@doi [\aj]
  {10.1088/0004-6256/136/4/1623}, \href
  {http://adsabs.harvard.edu/abs/2008AJ....136.1623C} {136, 1623}

\bibitem[\protect\citeauthoryear{{Dehnen}}{{Dehnen}}{1993}]{1993MNRAS.265..250D}
{Dehnen} W.,  1993, \mn@doi [\mnras] {10.1093/mnras/265.1.250}, \href
  {http://adsabs.harvard.edu/abs/1993MNRAS.265..250D} {265, 250}

\bibitem[\protect\citeauthoryear{{Domainko} et~al.,}{{Domainko}
  et~al.}{2006}]{2006A&A...452..795D}
{Domainko} W.,  et~al., 2006, \mn@doi [\aap] {10.1051/0004-6361:20053921},
  \href {http://adsabs.harvard.edu/abs/2006A%26A...452..795D} {452, 795}

\bibitem[\protect\citeauthoryear{{Dressler}}{{Dressler}}{1980}]{1980ApJ...236..351D}
{Dressler} A.,  1980, \mn@doi [\apj] {10.1086/157753}, \href
  {http://adsabs.harvard.edu/abs/1980ApJ...236..351D} {236, 351}

\bibitem[\protect\citeauthoryear{{Duffy}, {Schaye}, {Kay}  \& {Dalla
  Vecchia}}{{Duffy} et~al.}{2008}]{2008MNRAS.390L..64D}
{Duffy} A.~R.,  {Schaye} J.,  {Kay} S.~T.,   {Dalla Vecchia} C.,  2008, \mn@doi
  [\mnras] {10.1111/j.1745-3933.2008.00537.x}, \href
  {http://adsabs.harvard.edu/abs/2008MNRAS.390L..64D} {390, L64}

\bibitem[\protect\citeauthoryear{{Ebeling}, {Stephenson}  \& {Edge}}{{Ebeling}
  et~al.}{2014}]{2014ApJ...781L..40E}
{Ebeling} H.,  {Stephenson} L.~N.,   {Edge} A.~C.,  2014, \mn@doi [\apjl]
  {10.1088/2041-8205/781/2/L40}, \href
  {http://adsabs.harvard.edu/abs/2014ApJ...781L..40E} {781, L40}

\bibitem[\protect\citeauthoryear{{Farouki} \& {Shapiro}}{{Farouki} \&
  {Shapiro}}{1980}]{1980ApJ...241..928F}
{Farouki} R.,  {Shapiro} S.~L.,  1980, \mn@doi [\apj] {10.1086/158408}, \href
  {http://adsabs.harvard.edu/abs/1980ApJ...241..928F} {241, 928}

\bibitem[\protect\citeauthoryear{{Ferri{\`e}re}}{{Ferri{\`e}re}}{2001}]{2001RvMP...73.1031F}
{Ferri{\`e}re} K.~M.,  2001, \mn@doi [Reviews of Modern Physics]
  {10.1103/RevModPhys.73.1031}, \href
  {http://adsabs.harvard.edu/abs/2001RvMP...73.1031F} {73, 1031}

\bibitem[\protect\citeauthoryear{{Fujita} \& {Nagashima}}{{Fujita} \&
  {Nagashima}}{1999}]{1999ApJ...516..619F}
{Fujita} Y.,  {Nagashima} M.,  1999, \mn@doi [\apj] {10.1086/307139}, \href
  {http://adsabs.harvard.edu/abs/1999ApJ...516..619F} {516, 619}

\bibitem[\protect\citeauthoryear{{Gill}, {Knebe}  \& {Gibson}}{{Gill}
  et~al.}{2005}]{2005MNRAS.356.1327G}
{Gill} S.~P.~D.,  {Knebe} A.,   {Gibson} B.~K.,  2005, \mn@doi [\mnras]
  {10.1111/j.1365-2966.2004.08562.x}, \href
  {http://adsabs.harvard.edu/abs/2005MNRAS.356.1327G} {356, 1327}

\bibitem[\protect\citeauthoryear{{Gunn} \& {Gott}}{{Gunn} \&
  {Gott}}{1972}]{1972ApJ...176....1G}
{Gunn} J.~E.,  {Gott} III J.~R.,  1972, \mn@doi [\apj] {10.1086/151605}, \href
  {http://adsabs.harvard.edu/abs/1972ApJ...176....1G} {176, 1}

\bibitem[\protect\citeauthoryear{{Hernquist}}{{Hernquist}}{1990}]{1990ApJ...356..359H}
{Hernquist} L.,  1990, \mn@doi [\apj] {10.1086/168845}, \href
  {http://adsabs.harvard.edu/abs/1990ApJ...356..359H} {356, 359}

\bibitem[\protect\citeauthoryear{{Johnson}, {Harrison}, {Swinbank}, {Bower},
  {Smail}, {Koyama}  \& {Geach}}{{Johnson} et~al.}{2016}]{2016MNRAS.460.1059J}
{Johnson} H.~L.,  {Harrison} C.~M.,  {Swinbank} A.~M.,  {Bower} R.~G.,  {Smail}
  I.,  {Koyama} Y.,   {Geach} J.~E.,  2016, \mn@doi [\mnras]
  {10.1093/mnras/stw1030}, \href
  {http://adsabs.harvard.edu/abs/2016MNRAS.460.1059J} {460, 1059}

\bibitem[\protect\citeauthoryear{{Kauffmann}, {White}, {Heckman}, {M{\'e}nard},
  {Brinchmann}, {Charlot}, {Tremonti}  \& {Brinkmann}}{{Kauffmann}
  et~al.}{2004}]{2004MNRAS.353..713K}
{Kauffmann} G.,  {White} S.~D.~M.,  {Heckman} T.~M.,  {M{\'e}nard} B.,
  {Brinchmann} J.,  {Charlot} S.,  {Tremonti} C.,   {Brinkmann} J.,  2004,
  \mn@doi [\mnras] {10.1111/j.1365-2966.2004.08117.x}, \href
  {http://adsabs.harvard.edu/abs/2004MNRAS.353..713K} {353, 713}

\bibitem[\protect\citeauthoryear{{Kenney}, {Abramson}  \&
  {Bravo-Alfaro}}{{Kenney} et~al.}{2015}]{2015AJ....150...59K}
{Kenney} J.~D.~P.,  {Abramson} A.,   {Bravo-Alfaro} H.,  2015, \mn@doi [\aj]
  {10.1088/0004-6256/150/2/59}, \href
  {http://adsabs.harvard.edu/abs/2015AJ....150...59K} {150, 59}

\bibitem[\protect\citeauthoryear{{Koopmann} \& {Kenney}}{{Koopmann} \&
  {Kenney}}{2004}]{2004ApJ...613..866K}
{Koopmann} R.~A.,  {Kenney} J.~D.~P.,  2004, \mn@doi [\apj] {10.1086/423191},
  \href {http://adsabs.harvard.edu/abs/2004ApJ...613..866K} {613, 866}

\bibitem[\protect\citeauthoryear{{Kronberger}, {Kapferer}, {Ferrari},
  {Unterguggenberger}  \& {Schindler}}{{Kronberger}
  et~al.}{2008}]{2008A&A...481..337K}
{Kronberger} T.,  {Kapferer} W.,  {Ferrari} C.,  {Unterguggenberger} S.,
  {Schindler} S.,  2008, \mn@doi [\aap] {10.1051/0004-6361:20078904}, \href
  {http://adsabs.harvard.edu/abs/2008A%26A...481..337K} {481, 337}

\bibitem[\protect\citeauthoryear{{Krumholz} \& {Tan}}{{Krumholz} \&
  {Tan}}{2007}]{2007ApJ...654..304K}
{Krumholz} M.~R.,  {Tan} J.~C.,  2007, \mn@doi [\apj] {10.1086/509101}, \href
  {http://adsabs.harvard.edu/abs/2007ApJ...654..304K} {654, 304}

\bibitem[\protect\citeauthoryear{{Lagan{\'a}}, {Martinet}, {Durret}, {Lima
  Neto}, {Maughan}  \& {Zhang}}{{Lagan{\'a}}
  et~al.}{2013}]{2013A&A...555A..66L}
{Lagan{\'a}} T.~F.,  {Martinet} N.,  {Durret} F.,  {Lima Neto} G.~B.,
  {Maughan} B.,   {Zhang} Y.-Y.,  2013, \mn@doi [\aap]
  {10.1051/0004-6361/201220423}, \href
  {http://adsabs.harvard.edu/abs/2013A%26A...555A..66L} {555, A66}

\bibitem[\protect\citeauthoryear{{McPartland}, {Ebeling}, {Roediger}  \&
  {Blumenthal}}{{McPartland} et~al.}{2016}]{2016MNRAS.455.2994M}
{McPartland} C.,  {Ebeling} H.,  {Roediger} E.,   {Blumenthal} K.,  2016,
  \mn@doi [\mnras] {10.1093/mnras/stv2508}, \href
  {http://adsabs.harvard.edu/abs/2016MNRAS.455.2994M} {455, 2994}

\bibitem[\protect\citeauthoryear{{Muriel} \& {Coenda}}{{Muriel} \&
  {Coenda}}{2014}]{2014A&A...564A..85M}
{Muriel} H.,  {Coenda} V.,  2014, \mn@doi [\aap] {10.1051/0004-6361/201322033},
  \href {http://adsabs.harvard.edu/abs/2014A%26A...564A..85M} {564, A85}

\bibitem[\protect\citeauthoryear{{Navarro}, {Frenk}  \& {White}}{{Navarro}
  et~al.}{1996}]{1996ApJ...462..563N}
{Navarro} J.~F.,  {Frenk} C.~S.,   {White} S.~D.~M.,  1996, \mn@doi [\apj]
  {10.1086/177173}, \href {http://adsabs.harvard.edu/abs/1996ApJ...462..563N}
  {462, 563}

\bibitem[\protect\citeauthoryear{{Pimbblet}}{{Pimbblet}}{2011}]{2011MNRAS.411.2637P}
{Pimbblet} K.~A.,  2011, \mn@doi [\mnras] {10.1111/j.1365-2966.2010.17869.x},
  \href {http://adsabs.harvard.edu/abs/2011MNRAS.411.2637P} {411, 2637}

\bibitem[\protect\citeauthoryear{{Pimbblet}, {Smail}, {Edge}, {O'Hely}, {Couch}
   \& {Zabludoff}}{{Pimbblet} et~al.}{2006}]{2006MNRAS.366..645P}
{Pimbblet} K.~A.,  {Smail} I.,  {Edge} A.~C.,  {O'Hely} E.,  {Couch} W.~J.,
  {Zabludoff} A.~I.,  2006, \mn@doi [\mnras]
  {10.1111/j.1365-2966.2005.09892.x}, \href
  {http://adsabs.harvard.edu/abs/2006MNRAS.366..645P} {366, 645}

\bibitem[\protect\citeauthoryear{{Planck Collaboration} et~al.,}{{Planck
  Collaboration} et~al.}{2016}]{2016A&A...594A..13P}
{Planck Collaboration} et~al., 2016, \mn@doi [\aap]
  {10.1051/0004-6361/201525830}, \href
  {http://adsabs.harvard.edu/abs/2016A%26A...594A..13P} {594, A13}

\bibitem[\protect\citeauthoryear{{Quilis}, {Moore}  \& {Bower}}{{Quilis}
  et~al.}{2000}]{2000Sci...288.1617Q}
{Quilis} V.,  {Moore} B.,   {Bower} R.,  2000, \mn@doi [Science]
  {10.1126/science.288.5471.1617}, \href
  {http://adsabs.harvard.edu/abs/2000Sci...288.1617Q} {288, 1617}

\bibitem[\protect\citeauthoryear{{Robitaille} \& {Whitney}}{{Robitaille} \&
  {Whitney}}{2010}]{2010ApJ...710L..11R}
{Robitaille} T.~P.,  {Whitney} B.~A.,  2010, \mn@doi [\apjl]
  {10.1088/2041-8205/710/1/L11}, \href
  {http://adsabs.harvard.edu/abs/2010ApJ...710L..11R} {710, L11}

\bibitem[\protect\citeauthoryear{{Roediger} \& {Br{\"u}ggen}}{{Roediger} \&
  {Br{\"u}ggen}}{2006}]{2006MNRAS.369..567R}
{Roediger} E.,  {Br{\"u}ggen} M.,  2006, \mn@doi [\mnras]
  {10.1111/j.1365-2966.2006.10335.x}, \href
  {http://adsabs.harvard.edu/abs/2006MNRAS.369..567R} {369, 567}

\bibitem[\protect\citeauthoryear{{Roediger} \& {Br{\"u}ggen}}{{Roediger} \&
  {Br{\"u}ggen}}{2007}]{2007MNRAS.380.1399R}
{Roediger} E.,  {Br{\"u}ggen} M.,  2007, \mn@doi [\mnras]
  {10.1111/j.1365-2966.2007.12241.x}, \href
  {http://adsabs.harvard.edu/abs/2007MNRAS.380.1399R} {380, 1399}

\bibitem[\protect\citeauthoryear{{Roediger} \& {Br{\"u}ggen}}{{Roediger} \&
  {Br{\"u}ggen}}{2008}]{2008MNRAS.388L..89R}
{Roediger} E.,  {Br{\"u}ggen} M.,  2008, \mn@doi [\mnras]
  {10.1111/j.1745-3933.2008.00506.x}, \href
  {http://adsabs.harvard.edu/abs/2008MNRAS.388L..89R} {388, L89}

\bibitem[\protect\citeauthoryear{{Ruszkowski}, {Br{\"u}ggen}, {Lee}  \&
  {Shin}}{{Ruszkowski} et~al.}{2014}]{2014ApJ...784...75R}
{Ruszkowski} M.,  {Br{\"u}ggen} M.,  {Lee} D.,   {Shin} M.-S.,  2014, \mn@doi
  [\apj] {10.1088/0004-637X/784/1/75}, \href
  {http://adsabs.harvard.edu/abs/2014ApJ...784...75R} {784, 75}

\bibitem[\protect\citeauthoryear{{Schmidt}}{{Schmidt}}{1959}]{1959ApJ...129..243S}
{Schmidt} M.,  1959, \mn@doi [\apj] {10.1086/146614}, \href
  {http://adsabs.harvard.edu/abs/1959ApJ...129..243S} {129, 243}

\bibitem[\protect\citeauthoryear{{Springel}}{{Springel}}{2005}]{2005MNRAS.364.1105S}
{Springel} V.,  2005, \mn@doi [\mnras] {10.1111/j.1365-2966.2005.09655.x},
  \href {http://adsabs.harvard.edu/abs/2005MNRAS.364.1105S} {364, 1105}

\bibitem[\protect\citeauthoryear{{Springel}, {Di Matteo}  \&
  {Hernquist}}{{Springel} et~al.}{2005}]{2005MNRAS.361..776S}
{Springel} V.,  {Di Matteo} T.,   {Hernquist} L.,  2005, \mn@doi [\mnras]
  {10.1111/j.1365-2966.2005.09238.x}, \href
  {http://adsabs.harvard.edu/abs/2005MNRAS.361..776S} {361, 776}

\bibitem[\protect\citeauthoryear{{Steinhauser}, {Haider}, {Kapferer}  \&
  {Schindler}}{{Steinhauser} et~al.}{2012}]{2012A&A...544A..54S}
{Steinhauser} D.,  {Haider} M.,  {Kapferer} W.,   {Schindler} S.,  2012,
  \mn@doi [\aap] {10.1051/0004-6361/201118311}, \href
  {http://adsabs.harvard.edu/abs/2012A%26A...544A..54S} {544, A54}

\bibitem[\protect\citeauthoryear{{Steinhauser}, {Schindler}  \&
  {Springel}}{{Steinhauser} et~al.}{2016}]{2016A&A...591A..51S}
{Steinhauser} D.,  {Schindler} S.,   {Springel} V.,  2016, \mn@doi [\aap]
  {10.1051/0004-6361/201527705}, \href
  {http://adsabs.harvard.edu/abs/2016A%26A...591A..51S} {591, A51}

\bibitem[\protect\citeauthoryear{{Tecce}, {Cora}, {Tissera}, {Abadi}  \&
  {Lagos}}{{Tecce} et~al.}{2010}]{2010MNRAS.408.2008T}
{Tecce} T.~E.,  {Cora} S.~A.,  {Tissera} P.~B.,  {Abadi} M.~G.,   {Lagos}
  C.~D.~P.,  2010, \mn@doi [\mnras] {10.1111/j.1365-2966.2010.17262.x}, \href
  {http://adsabs.harvard.edu/abs/2010MNRAS.408.2008T} {408, 2008}

\bibitem[\protect\citeauthoryear{{Teyssier}}{{Teyssier}}{2002}]{2002A&A...385..337T}
{Teyssier} R.,  2002, \mn@doi [\aap] {10.1051/0004-6361:20011817}, \href
  {http://adsabs.harvard.edu/abs/2002A%26A...385..337T} {385, 337}

\bibitem[\protect\citeauthoryear{{Teyssier}, {Moore}, {Martizzi}, {Dubois}  \&
  {Mayer}}{{Teyssier} et~al.}{2011}]{2011MNRAS.414..195T}
{Teyssier} R.,  {Moore} B.,  {Martizzi} D.,  {Dubois} Y.,   {Mayer} L.,  2011,
  \mn@doi [\mnras] {10.1111/j.1365-2966.2011.18399.x}, \href
  {http://adsabs.harvard.edu/abs/2011MNRAS.414..195T} {414, 195}

\bibitem[\protect\citeauthoryear{{Teyssier}, {Pontzen}, {Dubois}  \&
  {Read}}{{Teyssier} et~al.}{2013}]{2013MNRAS.429.3068T}
{Teyssier} R.,  {Pontzen} A.,  {Dubois} Y.,   {Read} J.~I.,  2013, \mn@doi
  [\mnras] {10.1093/mnras/sts563}, \href
  {http://adsabs.harvard.edu/abs/2013MNRAS.429.3068T} {429, 3068}

\bibitem[\protect\citeauthoryear{{Tonnesen} \& {Bryan}}{{Tonnesen} \&
  {Bryan}}{2008}]{2008ApJ...684L...9T}
{Tonnesen} S.,  {Bryan} G.~L.,  2008, \mn@doi [\apjl] {10.1086/592066}, \href
  {http://adsabs.harvard.edu/abs/2008ApJ...684L...9T} {684, L9}

\bibitem[\protect\citeauthoryear{{Tonnesen} \& {Stone}}{{Tonnesen} \&
  {Stone}}{2014}]{2014ApJ...795..148T}
{Tonnesen} S.,  {Stone} J.,  2014, \mn@doi [\apj]
  {10.1088/0004-637X/795/2/148}, \href
  {http://adsabs.harvard.edu/abs/2014ApJ...795..148T} {795, 148}

\bibitem[\protect\citeauthoryear{{Tonnesen}, {Bryan}  \& {van
  Gorkom}}{{Tonnesen} et~al.}{2007}]{2007ApJ...671.1434T}
{Tonnesen} S.,  {Bryan} G.~L.,   {van Gorkom} J.~H.,  2007, \mn@doi [\apj]
  {10.1086/523034}, \href {http://adsabs.harvard.edu/abs/2007ApJ...671.1434T}
  {671, 1434}

\bibitem[\protect\citeauthoryear{{Turk}, {Smith}, {Oishi}, {Skory}, {Skillman},
  {Abel}  \& {Norman}}{{Turk} et~al.}{2011}]{2011ApJS..192....9T}
{Turk} M.~J.,  {Smith} B.~D.,  {Oishi} J.~S.,  {Skory} S.,  {Skillman} S.~W.,
  {Abel} T.,   {Norman} M.~L.,  2011, \mn@doi [\apjs]
  {10.1088/0067-0049/192/1/9}, \href
  {http://adsabs.harvard.edu/abs/2011ApJS..192....9T} {192, 9}

\bibitem[\protect\citeauthoryear{{van der Wel}, {Bell}, {Holden}, {Skibba}  \&
  {Rix}}{{van der Wel} et~al.}{2010}]{2010ApJ...714.1779V}
{van der Wel} A.,  {Bell} E.~F.,  {Holden} B.~P.,  {Skibba} R.~A.,   {Rix}
  H.-W.,  2010, \mn@doi [\apj] {10.1088/0004-637X/714/2/1779}, \href
  {http://adsabs.harvard.edu/abs/2010ApJ...714.1779V} {714, 1779}

\makeatother
\end{thebibliography}

\bsp	
\label{lastpage}
\end{document}